# Rethinking the Unpretentious U-net for Medical Ultrasound Image Segmentation

Gongping Chen, Lei Li, Jianxun Zhang and Yu Dai

**Abstract**—Breast tumor segmentation is one of the key steps that helps us characterize and localize tumor regions. However, variable tumor morphology, blurred boundary, and similar intensity distributions bring challenges for accurate segmentation of breast tumors. Recently, many U-net variants have been proposed and widely used for breast tumors segmentation. However, these architectures suffer from two limitations: (1) Ignoring the characterize ability of the benchmark networks, and (2) Introducing extra complex operations increases the difficulty of understanding and reproducing the network. To alleviate these challenges, this paper proposes a simple yet powerful nested U-net (NU-net) for accurate segmentation of breast tumors. The key idea is to utilize U-Nets with different depths and shared weights to achieve robust characterization of breast tumors. NU-net mainly has the following advantages: (1) Improving network adaptability and robustness to breast tumors with different scales, (2) This method is easy to reproduce and execute, and (3) The extra operations increase network parameters without significantly increasing computational cost. Extensive experimental results with twelve state-of-the-art segmentation methods on three public breast ultrasound datasets demonstrate that NU-net has more competitive segmentation performance on breast tumors. Furthermore, the robustness of NU-net is further illustrated on the segmentation of renal ultrasound images. The source code is publicly available on https://github.com/CGPxy/NU-net.

—————————— ◆ ——————————

## 1 INTRODUCTION

Breast cancer is one of the most dreaded cancers in women and it seriously threatens women's health [1]. At present, ultrasound imaging is widely used in clinical pre-screening due to its advantages of flexibility, harmlessness and low cost. Segmenting the lesion region from breast ultrasound images is essential for tumor diagnosis and postoperative follow-up [2]–[4]. However, complex ultrasound pattern, similar intensity distribution, variable tumor morphology, and blurred boundary bring great challenges for automatic segmentation of breast tumors [5], as shown in Fig. 1.

Due to the powerful nonlinear representation ability, convolutional neural networks (CNNs) have been successfully and widely used in medical image segmentation [6]–[9] or natural image segmentation [10]–[12]. Among them, U-net [8] is one of the most successful network architectures in medical image segmentation. It uses skip-connections to fuse shallow, low-level, fine-grained feature maps in the encoder with deep, semantic, and coarse-grained feature maps in the decoding to improve the segmentation accuracy of the network [13]. Motivated by this, Almajalid et al. [14] used U-net to segment enhanced and denoised breast ultrasound images for the first time. However, these preprocessing operations weaken the spatial structure of the object and smooth the edges. Moreover, the original U-net cannot cope with the perturbations of surrounding tissue and tumor morphology, making it difficult to learn robust representations of breast tumors from complex ultrasound images, as shown in Fig. 1.

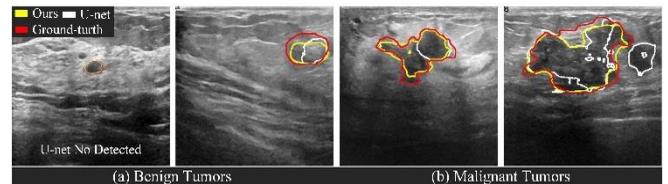

Fig. 1. The segmentation results of U-net and our method on breast tumors. It can be seen from these images that variable morphology blurred boundary and similar surrounding issue severely affect the segmentation accuracy of breast tumors, especially for small and malignant tumors.

To further improve the segmentation accuracy of breast tumors, many variant networks based on encoder-decoder architectures have been developed to segment breast tumors [15]–[19]. These variant networks can be roughly classified into four types: multi-scale U-net [20]–[23], attention-optimized U-net [24]–[28], deep-supervised U-net [29]–[31], and multi-module hybrid U-net [32]–[36], as shown in Fig. 2. Existing work has indicated that introducing different strategies (such as residual learning, attention module, multi-scale, deep supervision) can improve the segmentation performance of U-net on breast tumors [19]. However, these variant architectures also suffer from two distinct limitations.

**(A) Robustness.** Although the introduction of these additional operations improved the segmentation accuracy of breast tumors, they did not adequately dissect the characterize ability of the benchmark network. Therefore,

- G.-P. Chen, Y. Dai and J.-X. Zhang are with the Medical & Robotics Lab, College of Artifical Intelligence, Nankai University, Tianjin, China. (Email: cgp110@mail.nankai.edu.cn, daiyu@nankai.edu.cn, zhangjx@nankai.edu.cn
- Lei Li is with the Institute of Biomedical Engineering, University of Oxford London, UK. (Email: lei.li@eng.ox.ac.uk)
- Corresponding author: Y. Dai

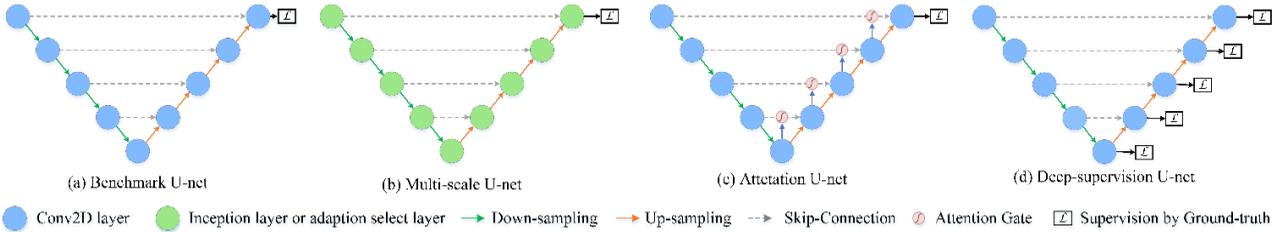

Fig. 2. The rough illustration of U-net and its variant networks.

we can reasonably assume that introducing additional operations to the benchmark U-net with good representation ability can achieve more robust segmentation performance.

**(B) Complexities.** These extra operations inevitably increase the difficulty of understanding and reproducing the network, which is bound to hinder the wide application of the network. Therefore, we need the proposed network to be easily reproduced and implemented.

To address the aforementioned limitations, we develop a simple nested U-net (NU-net) to improve the segmentation accuracy of breast tumors. Specifically, we first utilize the deeper U-net (fifteen layers) as the baseline network. Then, the developed multi-out U-net is embedded as the bones between encoder and decoder. Finally, three short-connections based on multi-step down-sampling are used to enhance the correlation of long-range information of encoded features. Compared with the variant networks of U-net, NU-net mainly has the following advantages:

- First, NU-net has no complex components without significantly increasing the computational cost and is easy to reproduce and execute.
- Second, simply increasing the depth of the baseline U-net significantly improves the breast tumors segmentation accuracy and outperforms many variant networks.
- Third, the nested multi-out U-nets further enhances the correlation between fine-grained and semantic features while refining the encoded feature maps. Furthermore, it can improve the adaptability of the network to breast tumors with different scales.
- Moreover, extensive experimental results on public breast ultrasound datasets show that NU-net has better robustness in breast tumors segmentation.

## 2. RELATED WORK

### 2.1 Multi-Scale U-net

A common way to build the multi-scale U-shaped network is to use inception layers with different kernel sizes [10] as the convolutional layers, such as STAN [20] and ARFNet [22]. However, Li et al. [37] pointed out that this artificially designed multi-scale convolutional layers cannot adaptively extract objective information under different scale receptive fields. To adaptively capture the feature information of breast tumors under different receptive fields, Byra et al. [21] developed a selective kernel U-net (SKU-net) to segment breast tumors. Chen et al. [23] pointed out that there is an obvious limitation of SKU-net, which only considers feature selection in the channel dimension. To overcome this limitation, an adaptive attention U-net (AAU-net) is proposed to adaptively select breast tumors features under different receptive fields from channel and spatial dimensions [23]. AAU-net can select more representative breast tumor features, but the operation in spatial and channel dimensions is computationally expensive. In addition, fully adapting to the variable tumor morphology remains a challenge.

### 2.2 Attention Optimized U-net

With the advent of attention U-net (Att U-net) [38], many U-shaped networks integrated with attention modules have also been developed to segment breast tumors. Yan et al. [27] proposed an attention enhanced U-net (AE U-net) with hybrid dilated convolution based on attention U-net to segment the breast tumors from ultrasound images. However, the use of dilated convolutions on deeper convolutional layers cannot capture sufficient contextual information [39]. In addition, its segmentation performance on breast lesions is limited by Att U-net. Lee et al. [25] improved the segmentation performance of U-net for breast tumors by introducing the channel attention module with multiscale grid average pooling. Subsequently, Lei et al. [24] designed a more complex network to segment breast tumors by integrating spatial attention, channel attention and non-local modules. However, similar surrounding tissue can interfere with the performance of non-local operations. Chen et al. [28] designed a bidirectional aware-guided network using bidirectional calibration of global features and local features to improve the robustness of the network. Although the introduction of different attention modules improves the segmentation performance of U-net, they still suffer from various disturbance factors.

### 2.3 Deep-Supervised U-net

Unlike multi-scale U-net and attention-optimized U-net, deep-supervised U-net constrains each stage of the network to make the learned characterizes closer to the ground-truth objective. Wang et al. [29] used deep supervision strategy constraints on the feature maps captured at each stage of U-net to segment breast lesions. However, excessively introducing deep supervision constraints will not improve the performance of the network and will increase the network parameters. Qu et al. [30] improved the segmentation accuracy of U-net for breast tumors by adding deep supervision in the decoding stage. The introduction of the deep supervision mechanism can guide the network to learn to predict segmentation results scale-by-scale, and this strategy can improve the segmentation accuracy of breast tumors to a certain extent.

TABLE 1
THE SUMMARY OF U-NET VARIANT NETWORKS FOR SEGMENTATION OF BREAST ULTRASOUND IMAGES.

| Method | Year | Dataset | Main External Operations |
|---|---|---|---|
| PPU-net [14] | 2018 | Private Dataset | Contrast enhancement, Despeckling, Denoising |
| STAN [20] | 2020 | Dataset B [2]+BUSIS [40] | Multi-scale |
| ARFNet [22] | 2022 | Dataset B [2] | Selecting multi-scale receptive fileds (Channel) |
| SKU-net [21] | 2020 | Dataset B [2]+BUSI [41]+OASBUD [42] | Selecting multi-scale receptive fileds (Channel) |
| AAU-net [23] | 2022 | Dataset B [2]+BUSI [41] +STU [32] | Selecting multi-scale receptive fileds (Channel and Spatial) |
| Att U-net [38] | 2017 | / | Spatial attentaton |
| MSGRAP [25] | 2020 | Dataset B [2] | Spatial attentation |
| SCANet [24] | 2020 | Private Dataset | Spatial and channel attention, Non-local block, Residual-learning |
| AE U-net [27] | 2020 | Private Dataset | Channel attention, Hybrid dilated-convolution |
| BAGNet [28] | 2022 | BUSI [41] | Spatial attention, Global attention |
| Wang et al. [29] | 2020 | Private Dataset | Deep-supervision |
| Qu et al. [30] | 2020 | Dataset B [2]+ Private Dataset | Deep-supervision, Residual-learning, Global attention |
| MADU-net [33] | 2019 | Dataset B [2] | Multi-scale, Spatial attention, Deep-supervision |
| MALF U-net [34] | 2021 | Private Dataset | Spatial attention, Residual-learning |
| RDAU-net [32] | 2019 | Dataset B [2] +STU [32] | Spatial attention, Residual-learning, Hybrid dilated-convolution |
| RCA-IUnet [35] | 2022 | BUSIS [40] +BUSI [41] | Spatial attention, Residual-learning, Multi-scale |
| IUnet [36] | 2020 | Private Dataset | Multi-scale, Deep-supervision |
| Ours | 2022 | Dataset B [2] +BUSI [41] +STU [32] | Multi-out U-net, Short-connection |

## 2.4 Multi-Module Hybrid U-net

To enable U-net to achieve better segmentation performance on breast tumors, residual learning, multi-scale, attention and deep supervision mechanisms are jointly introduced into U-net. Abraham et al. [33] constructed a new U-shaped network (MADU-net) to segment breast ultrasound images by introducing a multi-scale image input pyramid and deep supervision mechanism into the Att U-net. Multi-image inputs can provide more fine-grained feature maps, but introducing too many low-level feature maps will affect the characterize high-level semantic features and reduce the performance of the network. Tong et al. [34] utilized residual convolution blocks to replace convolution blocks of Att U-net to segment breast tumors. Subsequently, Zhuang et al. [32] introduced dilated convolution layers based on the work of Tong et al. [34] to capture features under different receptive fields. However, the objective feature captured on deeper convolutional layers lack contextual information [39]. In addition, these residual operations are often performed on feature maps with the same scale. To better capture the multi-scale information of the breast tumors, Punn et al. [35] replaced the convolution blocks of Att U-net with the residual blocks constructed by the inception convolution layer. Similarly, Wang et al. [36] improved the segmentation accuracy of U-net on breast tumors by integrating the inception convolution layer and the deep supervision strategy. However, Li et al. [37] pointed out that the multi-scale information captured by the inception convolution layer is more dependent on the artificially set convolution kernel size and cannot adaptively capture the multi-scale information of breast tumors. A summary of the existing U-shaped network for segmenting breast tumors is provided in TABLE 1.

## 3 METHOD

This section begins with a statement of the overview. Then, the architecture and implementation technical details of the segmentation network are introduced.

### 3.1 Overview

Through the systematic analysis of U-net and its variant networks on breast ultrasound images segmentation, we discover some of their limitations: (1) Tending to use shallower U-nets. (2) Introducing complex extra operations. (3) Reproducing and applying is inconvenient. To alleviate these limitations and further improve the segmentation accuracy of breast tumors, we propose a simple nested U-net (NU-net), as shown in Fig. 3. NU-net can be roughly regarded as a combination of seven U-nets with different depths and shared weights. Therefore, it is very easy to understand and reproduce. Specifically, we first utilize the deeper U-net (fifteen layers) as the backbone network to extract more sufficient breast tumors features. Then, the developed multi-out U-nets (MOU) is embedded as the bones between encoder and decoder. The nested multi-out U-nets can further enhance the correlation between fine-grained and semantic features while refining the encoded feature maps. Finally, the short-connections based on multi-step down-sampling (MDSC) are used to enhance the correlation of long-range information of encoded features.

### 3.2 Deeper Backbone U-net

U-net and its variant networks are widely used in breast lesions segmentation. However, the depth of these variant networks tends to be 9 or shallower. Existing research also shows that increasing the depth of the network can improve the generalization ability and robustness of the network [43]. Therefore, these shallow network architect-tures are often unable to adequately extract and characterize the spatial and location information of breast tumors for complex ultrasound images (such as: low-quality, variable morphology, similar surrounding tissue and blurred boundary). To capture sufficient and effective objective features from complex breast ultrasound images, we use U-net with a depth of fifteen as the baseline network for our method, See Section 5.1.

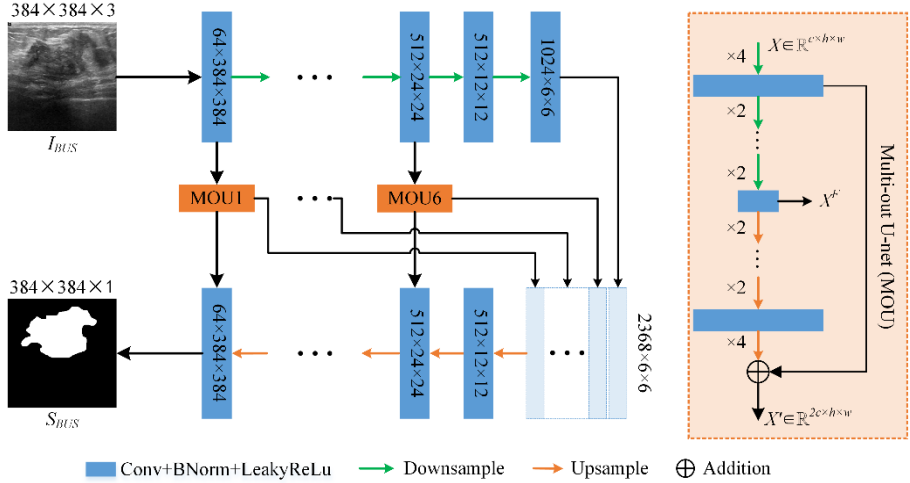

Fig. 3. Illustration of the NU-net, which is mainly composed of the deeper backbone U-net (fifteen layers) and the multi-out U-net (MOU). NU-net can be roughly regarded as a combination of seven U-nets with different depths and shared weights.

### 3.3 Multi-Out U-net (MOU)

It is well known that the roles of encoder and decoder in U-shaped network are to extract and reconstruct objective features, respectively. Therefore, the characterize ability of the encoder and decoder to the objective directly affects the performance of the network. However, how to ensure that the feature maps captured by each convolutional layer of the encoder are effective and fully utilized is a challenge. In this paper, we design a multi-out U-net to refine each set of feature maps in the encoding stage. As shown in Fig. 3, NU-net contains of six MOU moudles, and their depths are 11, 9, 7, 5, 3, 1, respectively. The filter size of all convolution layers in each MOU module is the same, which is the number of channels of the input feature maps. The MOU module mainly has three key functions: (1) The refinement of each set of encoded feature maps through the MOU module can help the network to better handle more complex breast ultrasound images. (2) The intermediate layer output of each MOU module is combined with the output of the backbone network encoder and fed into the backbone network decoder. Feature maps $X^f$ captured by the MOU modules with different depths can provide sufficient feature information for the reconstruction process of the decoder. (3) The use of MOU modules further enhances the correlation between encoding features and decoding features at the same scale. Through the embedding of the MOU module, we can roughly regard NU-net as the combination of seven U-nets with different depths and shared weights.

### 3.4 Multi-Step Down-Sampling Short-Connection

Simonyan et al. [43] pointed out that deep convolutional neural networks are prone to network degradation. The common approach to alleviate network degradation is to utilize short-connections to map the fine-grained features to the high-level features [44], as shown in Fig. 4(a) and Fig. 4(b). However, these short-connections are often implemented on feature maps with the same or adjacent scales, which cannot fully compensate for the loss of long-distance information. To further improve the correlation of distant feature maps, we utilize multi-step down-sampling as the short-connection operation, as shown in Fig. 4(c). The multi-step down-sampling short-connection consists of a pooling layer with pool-size $4 \times 4$ and a convolutional layer with filter size 32 accompanied by a batch normalization layer and activation layer. The encoder of our method contains three multi-step down-sampling short connections, as shown in Fig. 4(d). The combination of multi-out U-nets and multi-step down-sampling short-connections can alleviate the interference of breast tumor shape and scale on segmentation results, and improve the adaptability of the network to inputs with different scales.

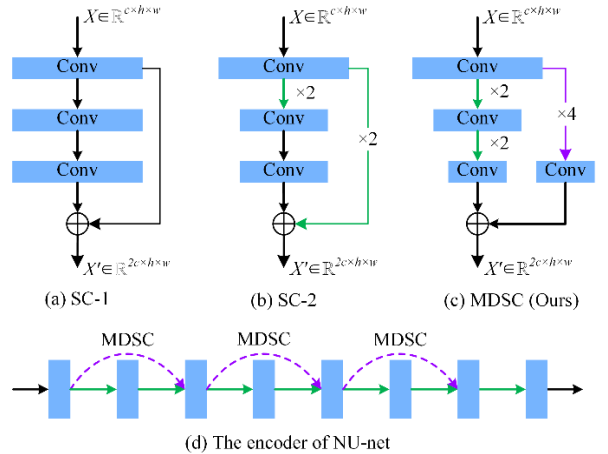

Fig. 4. Illustration of the different short-connections (SC).

## 4 MATERIALS AND EXPERIMENTS

In this section, we first introduce three public breast ul-trasound datasets. Then, the experimental detail for network training is described in detail. Finally, five evaluation metrics commonly used to evaluate network performance are introduced.

TABLE 2
Ablation Study of Different Network Components. We Perform Four-Fold Cross-Validation on BUSI and Dataset B, Respectively. The Best Results are Marked with Bold Texts.

| Methods | BUSI | | | | | Dataset B | | | | |
|---|---|---|---|---|---|---|---|---|---|---|
| | Jaccard | Precision | Recall | Specificity | Dice | Jaccard | Precision | Recall | Specificity | Dice |
| U-net | 60.70±2.36 | 71.88±2.41 | 76.30±2.48 | 96.18±0.55 | 70.10±2.20 | 58.44±4.26 | 70.27±6.11 | 75.32±2.85 | 98.44±0.40 | 68.20±4.23 |
| Deeper U-net | 68.91±1.88 | 78.70±2.79 | 81.50±2.31 | 97.36±0.51 | 77.31±1.95 | 67.86±1.87 | 76.59±2.76 | 80.95±2.31 | 98.65±0.55 | 76.48±1.77 |
| Deeper U-net + MOU | 70.26±1.45 | 79.37±1.27 | 82.34±1.99 | 97.39±0.59 | 78.58±1.41 | 69.76±1.37 | 79.12±1.75 | 82.95±1.56 | 98.84±0.33 | 78.70±0.87 |
| Deeper U-net + MOU + MDSC | **70.35±1.54** | **79.56±1.17** | **82.46±1.02** | **97.48±0.49** | **78.62±1.38** | **72.03±0.82** | **81.49±0.44** | **84.13±1.73** | **98.91±0.25** | **80.80±0.57** |

### 4.1 Dataset Description

In this paper, three widely used public breast ultrasound datasets are used to evaluate the segmentation network performance. The first breast ultrasound dataset (denotes as BUSI) is constructed by Al-Dhabyani et al. [41] contains 780 images of 600 female patients. Among them, 133 normal cases, 437 benign tumors, and 210 malignant tumors. These images acquired by two types of ultrasound equipment (LOGIQ E9 ultrasound and LOGIQ E9 Agile ultrasound system) in the Baheya Hospital. The second breast ultrasound dataset used in this paper named Dataset B is collected by Yap et al. [2]. Dataset B contains 163 images collected by Siemens ACUSON Sequoia C512 system. The benign and malignant cases in the Dataset B are 110 and 53, respectively. The third public breast ultrasound dataset is the STU provided by Zhuang et al. [32]. The STU contains 42 breast ultrasound images acquired by Shantou University using the GE Voluson E10 ultrasonic diagnostic system. The STU dataset does not distinguish benign and malignant cases. Since the STU dataset contains too few images, it is only used as external test data to evaluate the generalization performance of the segmentation network.

### 4.2 Evaluation Metrics

In this paper, five widely-used segmentation metrics are used to quantitatively evaluate the performance of different methods on breast ultrasound images segmentation. They are Jaccard, Precision, Recall, Specificity and Dice [23]. The higher the value of the five indicators of Jaccard, Precision, Recall, Specificity and Dice, the better the segmentation result of the network.

### 4.3 Experiment Details

To fully verify the effectiveness and robustness of our method, we use three breast ultrasound datasets to conduct extensive experiments, such as ablation study, comparison with state-of-the-art segmentation methods, and robustness analysis. In the ablation study, we perform four-fold cross-validation on BUSI and Dataset B, respectively. Similarly, we execute four-fold cross-validation on BUSI and Dataset B in our comparative experiments. In the robustness analysis experiments, we implement four-fold and three-fold cross-validation for benign and ma-lignant tumors in BUSI, respectively. Finally, we conduct external validation experiments on the already trained segmentation networks.

The loss function used for network training is binary cross-entropy. We utilize the Adam optimizer to train our network and its hyper parameters are set to the default values, where the initial learning rate is 0.001. Multiple cross-validation shows that the best segmentation performance is obtained when epoch size and batch size are set to 50 and 12, respectively. The development environment is TensorFlow 2.6.0, Python 3.6 and two NVIDIA RTX 3090 GPU.

## 5 Results

In this section, we first demonstrate the effectiveness of different components of the NU-net. Then, our method is compared with twelve state-of-the-art segmentation methods. Finally, we comprehensively evaluate the robustness of our approach.

### 5.1 Architecture Ablation

To demonstrate the effectiveness of the principal components of the NU-net, i.e., the deeper U-net, the multi-out U-net (MOU) and the multi-step down-sampling short-connection (MDSC), we conduct ablation experiments on BUSI and Dataset B. The baseline network is U-net, which includes four down-sampling and four up-sampling operations. In the ablation experiments, we implement four-fold cross-validation on BUSI and Dataset B, respectively. TABLE 2 illustrates the quantitative results of the architecture ablation study.

To obtain the best network depth, we compared U-nets with different depths. From Fig. 5, we can see that the segmentation performance on breast tumors can be further improved by increasing the depth of the benchmark U-net. This also indicates the necessity of increasing the network depth to cope with complex breast ultrasound images. As shown in Fig. 5, the best segmentation results can be achieved when the depth of U-net is 15 or 17. Therefore, we choose the 15-layer U-net with fewer network parameters as our baseline network. Comparing the results of U-net and Deeper U-net, it can be found that the increase of network depth significantly improves the segmentation performance of U-net on breast tumors. Through the comparison of Deeper U-net and 'Deeper U-net + MOU', we can see that embedding MOU modules can further improve the segmentation accuracy of breast tumors. This proves that refining the feature maps cap-tured by the encoder is beneficial to the network performance improvement. It can be concluded from the results of 'Deeper U-net + MOU' and 'Deeper U-net + MOU + MDSC' that the addition of multi-step down-sampling short-connections can further help the network to capture sufficient target features to improve the adaptability of the network. Through the above analysis, we can be concluded that these components designed in this paper are essential to improve the performance of U-net for breast tumors segmentation.

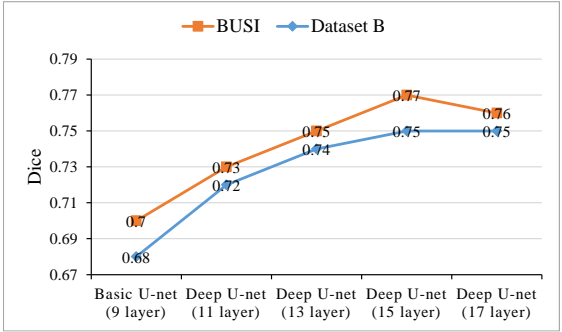

Fig. 5. The effect of increasing the depth of U-net on the segmentation results of BUSI and Dataset B.

TABLE 3
The Network Complexity of Different Components. The Definition of FLOPs Follow [45]

| | Params (M)/Multiple | GFLOPs/Multiple |
|---|---|---|
| U-net | 7.85/- | 62.82/- |
| Deeper U-net | 46.88/5.97 | 140.83/2.24 |
| Deeper U-net + MOU | 76.67/9.77 | 179.12/2.85 |
| Deeper U-net + MOU+ MDSC | 77.05/9.82 | 180.29/2.87 |

In addition, we analyze the complexity of different components in terms of network parameters and computational cost. TABLE 3 shows the network complexity of U-net after adding different components. Although increasing the network depth and MOU modules significantly increases network parameters, it does not significantly increase the computational cost. Furthermore, the addition of the MDSC has very little effect on increasing network parameters and computational cost. Overall, our method achieves more robust segmentation performance at the small computational cost.

## 5.2 Comparison with State-of-the-Art Methods

In this paper, our method is compared with twelve state-of-the-art deep learning methods for breast ultrasound images and medical images segmentation. Our comparative methods include U-net [8], Att U-net [38], U-net++ [13], U-net3+ [46], STAN [20], SKU-net [21], AAU-net [23], AE U-Net [27], RDAU-Net [32], MADU-net [33], scSEU-net [47], SegNet [11] and BASNet [31]. Among them, SegNet, U-net++, U-net3+, and BASNet are also U-shaped networks widely used in image segmentation. In the comparison experiment, we implement four-fold cross-validation on BUSI and Dataset B, respectively.

TABLE 4 shows the quantitative evaluation results of different methods on BUSI and Dataset B. The experimental results show that the proposed method achieves the most competitive segmentation results on BUSI and Dataset B. To further demonstrate the advantages of our method, we perform paired student's t-test with the second results, and the p-value ($p < 0.05$) indicates a significant difference between our method and the comparison methods.

## 5.3 Robustness Analysis

To demonstrate the robustness of NU-net, we first analyze the segmentation performance of NU-net on benign and malignant breast tumors. Then, we evaluate the segmentation performance of NU-net on different sites data.

TABLE 4
The Segmentation Results (mean ± std) of Different Competing Methods on BUSI and Dataset B. We Perform Four-Fold Cross-Validation on BUSI and Dataset B, Respectively. Asterisks Indicate That The Difference Between Our Method and The Competing Method is Significant Using a Paired Student's T-Test. (*: p < 0.05).

| Methods | BUSI | | | | | Dataset B | | | | |
|---|---|---|---|---|---|---|---|---|---|---|
| | Jaccard | Precision | Recall | Specificity | Dice | Jaccard | Precision | Recall | Specificity | Dice |
| U-net [8] | 60.70±2.36 | 71.88±2.41 | 76.30±2.48 | 96.18±0.55 | 70.10±2.20 | 58.44±4.26 | 70.27±6.11 | 75.32±2.85 | 98.44±0.40 | 68.20±4.23 |
| STAN [20] | 64.10±3.05 | 73.96±3.30 | 78.39±2.16 | 96.64±0.67 | 73.04±2.95 | 57.09±3.92 | 67.71±3.11 | 69.95±6.17 | 98.58±0.47 | 66.06±4.24 |
| Att U-net [38] | 57.09±1.22 | 78.78±4.67 | 66.97±4.08 | 96.87±0.83 | 67.99±1.18 | 59.93±4.53 | 70.40±6.05 | 76.15±4.21 | 98.43±0.33 | 69.30±4.07 |
| RDAU-net [32] | 63.75±3.36 | 71.25±4.11 | 78.90±1.35 | 96.63±0.76 | 71.94±3.46 | 58.17±4.91 | 70.49±4.26 | 73.55±5.28 | 98.37±0.39 | 68.22±4.94 |
| U-net++ [13] | 61.38±1.73 | 79.68±3.07 | 71.44±2.77 | 97.04±0.54 | 71.58±2.09 | 61.19±5.86 | 68.32±5.73 | 79.64±3.84 | 98.44±0.41 | 69.77±5.30 |
| MADU-net [33] | 61.62±2.69 | 73.77±2.90 | 76.87±2.58 | 96.40±0.62 | 71.35±2.67 | 63.09±3.04 | 73.70±5.08 | 79.24±1.72 | 98.61±0.36 | 72.32±3.14 |
| U-net3+ [46] | 63.03±2.79 | 71.89±3.28 | 79.58±2.48 | 96.19±0.68 | 71.85±2.73 | 65.63±5.26 | 73.50±6.21 | 80.29±3.93 | 98.60±0.36 | 73.98±4.72 |
| SegNet [11] | 67.31±1.87 | 76.09±2.00 | 79.85±1.03 | 96.99±0.53 | 75.64±1.80 | 62.83±2.20 | 71.72±1.70 | 80.15±3.90 | 98.59±0.30 | 72.16±1.52 |
| AE U-net [27] | 64.57±2.91 | 74.44±3.74 | 79.00±2.11 | 96.80±0.54 | 73.47±3.03 | 62.37±2.16 | 72.27±1.91 | 78.97±2.29 | 98.67±0.28 | 72.23±2.14 |
| scSEU-net [47] | 67.68±2.28 | 78.95±2.73 | 79.58±1.14 | 97.26±0.48 | 76.67±2.20 | 62.17±5.03 | 71.31±5.44 | 79.16±5.79 | 98.52±0.22 | 71.30±4.24 |
| BASNet [31] | 69.49±2.30* | 78.25±3.07 | 82.28±1.72* | 97.21±0.62 | 77.75±2.51* | 68.27±4.09 | 77.05±4.70 | 82.83±4.27* | 98.81±0.44 | 77.17±3.22 |
| SKU-net [21] | 68.10±1.63 | 78.62±1.66 | 79.53±1.93 | 97.33±0.45 | 76.92±1.57 | 64.25±4.01 | 75.27±6.70 | 79.36±2.50 | 98.68±0.39 | 73.53±4.05 |
| AAU-net [23] | 68.82±0.44 | **79.61±1.07** | 81.10±0.52 | **97.57±0.24** | 77.51±0.68 | 69.10±2.98* | 78.83±2.40* | 82.22±3.84 | 98.82±0.35* | 78.14±2.41* |
| Ours | **70.35±1.54** | 79.56±1.17 | **82.46±1.02** | 97.48±0.49 | **78.62±1.38** | **72.03±0.82** | **81.49±0.44** | **84.13±1.73** | **98.96±0.17** | **80.80±0.57** |

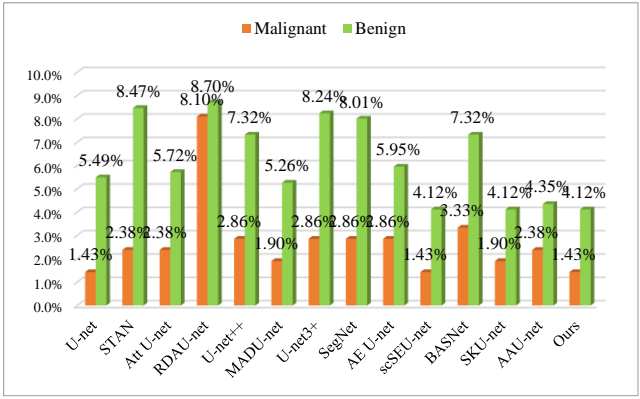

Fig. 6. The failure rates of different methods on the segmentation of benign and malignant tumors.

### 5.3.1 Robustness on Benign and Malignant Tumors

Benign tumors and malignant tumors have great differences such as morphology, boundary, intensity distribution, etc. To evaluate the robustness of the NU-net, we perform comparative analysis on malignant tumors and benign tumors, respectively. We conduct four-fold cross-validation on the benign tumors of BUSI and three-fold cross-validation on the malignant tumors of BUSI.

TABLE 5 show the segmentation results of the benign tumors and the malignant tumors. As shown in TABLE 5, our method still achieves the best performance on the segmentation of benign tumors and malignant tumors. Moreover, the p-value compared to the second results indicates that our method has a significant improvement in segmentation accuracy. Due to the complexity of ultrasound modalities, these variant networks still have serious missed detections and false detections on individual images, and even the segmentation fails. The proportion of segmentation failures of various methods on benign and malignant tumors is shown in Fig. 6. Our method achieves the smallest failure rate on the segmentation of benign and malignant tumors. The failure rates of our method on benign and malignant tumors are 1.43% and 4.12%, respectively.

### 5.3.2 External Validation

Due to the differences between different sites, there are large differences between the collected ultrasound images [26]. These differences can cause the model to perform well on the training dataset, but not perform well on the external dataset. To further evaluate the robustness of the proposed method in this paper, we used Dataset B and STU as external validation data to evaluate the adaptability of different methods to different site data. Specifically, we use Dataset B as the external verification data of BUSI, and STU as the external verification data of Dataset B. The segmentation results of various methods on the external validation datasets are presented in TABLE 6 and Fig. 7. Obviously, the proposed method achieves the best segmentation results in two external validation experiments.

## 6 APPLICATIONS

We further evaluate the effectiveness of our method on the renal ultrasound image segmentation task. To ensure the fairness of the experiments, we use the same experimental setup as recent work on renal segmentation to obtain the segmentation results of our method (refer to C-Net [48]). Similarly, we employ three-fold cross-validationto train our method. TABLE 7 shows the scores of various segmentation methods on the evaluation metrics for renal ultrasound images segmentation. Fig. 8 demonstrates the segmentation results of different methods on renal ultrasound images. Apparently, our method achieves the best segmentation performance, and outperforms all the compared methods. It can be seen from

TABLE 5
The Segmentation Results (mean ± std) of Benign and Malignant Lesions in BUSI by Different Methods. We Perform Four-Fold Cross-Validation on The Benign Lesions and Three-Fold Cross-Validation on The Malignant Lesions. Asterisks Indicate That The Difference Between Our Method and The Competing Method is Significant Using a Paired Student's T-Test. (*: $p < 0.05$).

| Methods | Benign tumors of BUSI | | | | | Malignant tumors of BUSI | | | | |
|---|---|---|---|---|---|---|---|---|---|---|
| | Jaccard | Precision | Recall | Specificity | Dice | Jaccard | Precision | Recall | Specificity | Dice |
| U-net [8] | 61.53±3.98 | 74.97±2.80 | 73.97±5.81 | 97.72±0.59 | 70.49±3.23 | 51.11±2.62 | 64.96±2.55 | 68.86±4.27 | 93.63±1.28 | 63.47±2.38 |
| STAN [20] | 64.20±2.73 | 73.37±3.70 | 77.91±1.27 | 97.65±0.52 | 71.98±2.98 | 51.01±2.38 | 62.74±3.72 | 70.96±6.75 | 93.69±1.56 | 62.50±1.97 |
| Att U-net [38] | 65.03±2.05 | 75.24±1.68 | 79.44±2.84 | 97.68±0.62 | 73.30±2.00 | 51.12±2.35 | 61.62±0.97 | 72.57±2.17 | 93.12±1.00 | 62.95±2.14 |
| RDAU-net [32] | 64.70±2.17 | 72.54±1.57 | 79.36±0.98 | 97.79±0.28 | 72.70±1.62 | 51.63±1.62 | 60.85±5.01 | 71.89±2.55 | 93.47±1.45 | 62.44±2.21 |
| U-net++ [13] | 68.25±2.75 | 75.93±3.66 | 81.58±1.09 | 97.74±0.62 | 75.56±2.79 | 54.03±3.03 | 65.50±2.94 | 73.43±2.10 | 93.73±1.31 | 65.52±2.75 |
| MADU-net [33] | 66.74±2.10 | 76.74±2.94 | 79.97±1.64 | 97.75±0.60 | 74.82±2.26 | 54.12±2.96 | 67.46±3.40 | 72.36±5.05 | 93.94±1.25 | 65.77±2.58 |
| U-net3+ [46] | 67.63±1.86 | 75.58±2.88 | 81.07±1.27 | 97.72±0.58 | 75.07±2.10 | 54.77±3.55 | 65.78±2.66 | 74.38±3.21 | 93.82±1.06 | 66.19±3.37 |
| SegNet [11] | 67.89±3.31 | 76.96±3.11 | 79.57±2.21 | 97.98±0.46 | 75.47±2.91 | 54.89±1.78 | 63.79±2.65 | 77.25±4.02* | 94.00±1.14 | 65.90±1.97 |
| AE U-net [27] | 67.89±1.96 | 77.17±3.63 | 80.54±1.25 | 97.95±0.60 | 75.77±1.82 | 55.38±1.77 | 67.87±3.81 | 72.88±3.47 | 94.43±1.33 | 66.50±1.52 |
| scSEU-net [47] | 71.33±2.45 | 80.76±2.69 | 81.58±2.45 | 98.20±1.47 | 78.97±2.44 | 56.21±2.16 | 67.75±4.77 | 73.99±3.22 | 94.29±1.13 | 67.39±1.96 |
| BASNet [31] | 70.55±2.38 | 78.00±2.83 | 82.25±1.17 | 98.13±0.54 | 77.78±2.29 | 58.94±1.71 | 68.68±2.15 | 76.30±3.36 | 94.79±0.90 | 69.27±1.51 |
| SKU-net [21] | 69.91±2.11 | 79.15±2.05 | 81.54±2.17 | 98.06±0.52 | 77.88±2.98 | 57.06±2.42 | 69.59±4.20 | 73.58±6.75 | 94.65±1.49 | 68.19±2.28 |
| AAU-net [23] | 73.33±2.09* | 82.70±2.90* | 83.14±0.87* | 98.39±0.47* | 80.88±2.06* | 60.60±1.70* | 72.62±3.13* | 76.13±5.66 | 95.11±1.27* | 71.54±1.74* |
| Ours | **74.34±2.83** | **82.91±2.42** | **85.56±3.59** | **98.43±0.40** | **81.43±2.85** | **61.37±0.96** | **72.88±1.90** | **77.41±2.99** | **95.15±1.14** | **72.15±0.70** |

TABLE 6
The Segmentation Results (mean ± std) of Different Competing Methods on the External Test Data STU and Dataset B. Asterisks Indicate That The Difference Between Our Method and The Competing Method is Significant Using a Paired Student's T-Test. (*: p < 0.05).

| Methods | Dataset B on BUSI | | | | | STU on Dataset B | | | | |
| --- | --- | --- | --- | --- | --- | --- | --- | --- | --- | --- |
| | Jaccard | Precision | Recall | Specificity | Dice | Jaccard | Precision | Recall | Specificity | Dice |
| U-net [8] | 42.72±4.65 | 63.44±5.55 | 54.48±6.86 | 98.17±0.70 | 53.31±4.31 | 58.90±3.75 | 66.27±5.46 | 86.88±1.60 | 94.54±0.74 | 71.41±3.67 |
| STAN [20] | 43.67±4.26 | 65.10±7.09 | 54.28±4.57 | 98.45±0.74 | 53.96±4.39 | 58.95±3.31 | 64.38±3.73 | 90.72±0.41 | 94.59±0.57 | 70.76±3.16 |
| Att U-net [38] | 38.48±4.19 | 47.72±5.37 | 58.38±4.06 | 97.38±0.71 | 47.36±5.57 | 52.65±2.29 | 59.26±2.86 | 86.35±1.29 | 93.41±0.41 | 65.19±2.73 |
| RDAU-net [32] | 40.42±3.36 | 57.51±7.36 | 51.14±2.01 | 98.25±0.74 | 49.26±4.05 | 60.11±3.02 | 63.49±2.86 | 91.37±1.07 | 94.70±0.45 | 72.40±2.74 |
| U-net++ [13] | 46.05±3.54 | 54.14±5.96 | 61.94±4.52 | 97.84±0.77 | 54.89±4.21 | 59.18±4.21 | 64.86±5.36 | 89.67±1.59 | 94.33±0.83 | 70.70±4.19 |
| MADU-net [33] | 44.19±2.29 | 56.40±3.38 | 62.11±2.28 | 97.81±0.65 | 54.34±3.01 | 58.11±3.24 | 66.05±3.69 | 86.17±0.99 | 94.35±0.47 | 70.32±3.01 |
| U-net3+ [46] | 42.70±2.93 | 51.18±3.96 | 62.20±2.18 | 97.64±0.72 | 51.03±3.04 | 61.51±1.78 | 67.12±2.22 | 89.96±1.01 | 94.49±0.31 | 72.96±1.69 |
| SegNet [11] | 42.56±8.30 | 63.39±10.94 | 51.59±9.73 | 98.29±0.87 | 51.61±9.02 | 62.70±3.09 | 66.57±3.05 | 91.36±0.38 | 95.04±0.49 | 73.50±3.62 |
| AE U-net [27] | 45.32±3.00 | 64.69±3.10 | 54.98±4.35 | 98.46±0.61 | 55.18±3.61 | 62.46±2.44 | 66.79±2.95 | 91.24±0.42 | 95.00±0.39 | 74.41±2.30 |
| scSEU-net [47] | 51.36±4.04* | 70.38±5.40 | 62.34±3.93* | 98.85±0.67 | 61.89±4.61* | 55.12±5.83 | 60.87±6.76 | 89.73±1.63 | 93.85±0.60 | 67.30±5.93 |
| BASNet [31] | 50.86±4.67 | 69.27±6.62 | 60.63±3.74 | 98.66±0.78 | 59.60±4.87 | 72.19±3.11* | 77.87±3.71* | 92.01±1.42 | 96.30±0.29* | 82.12±2.92* |
| SKU-net [21] | 47.25±3.99 | 76.23±4.27 | 53.27±4.86 | 98.88±0.66 | 57.79±4.45 | 66.94±3.15 | 71.99±4.08 | 91.44±1.08 | 95.40±0.51 | 78.29±3.05 |
| AAU-net [23] | 51.27±5.52 | 79.71±1.73* | 56.13±6.60 | 98.91±0.61* | 61.34±5.65 | 68.99±3.29 | 74.91±3.18 | 92.12±0.75* | 95.94±0.71 | 80.23±2.60 |
| Ours | **57.60±2.77** | **80.50±4.59** | **64.91±2.51** | **99.01±0.61** | **67.56±2.88** | **74.06±0.82** | **79.87±1.10** | **92.38±1.07** | **96.80±0.24** | **84.10±0.76** |

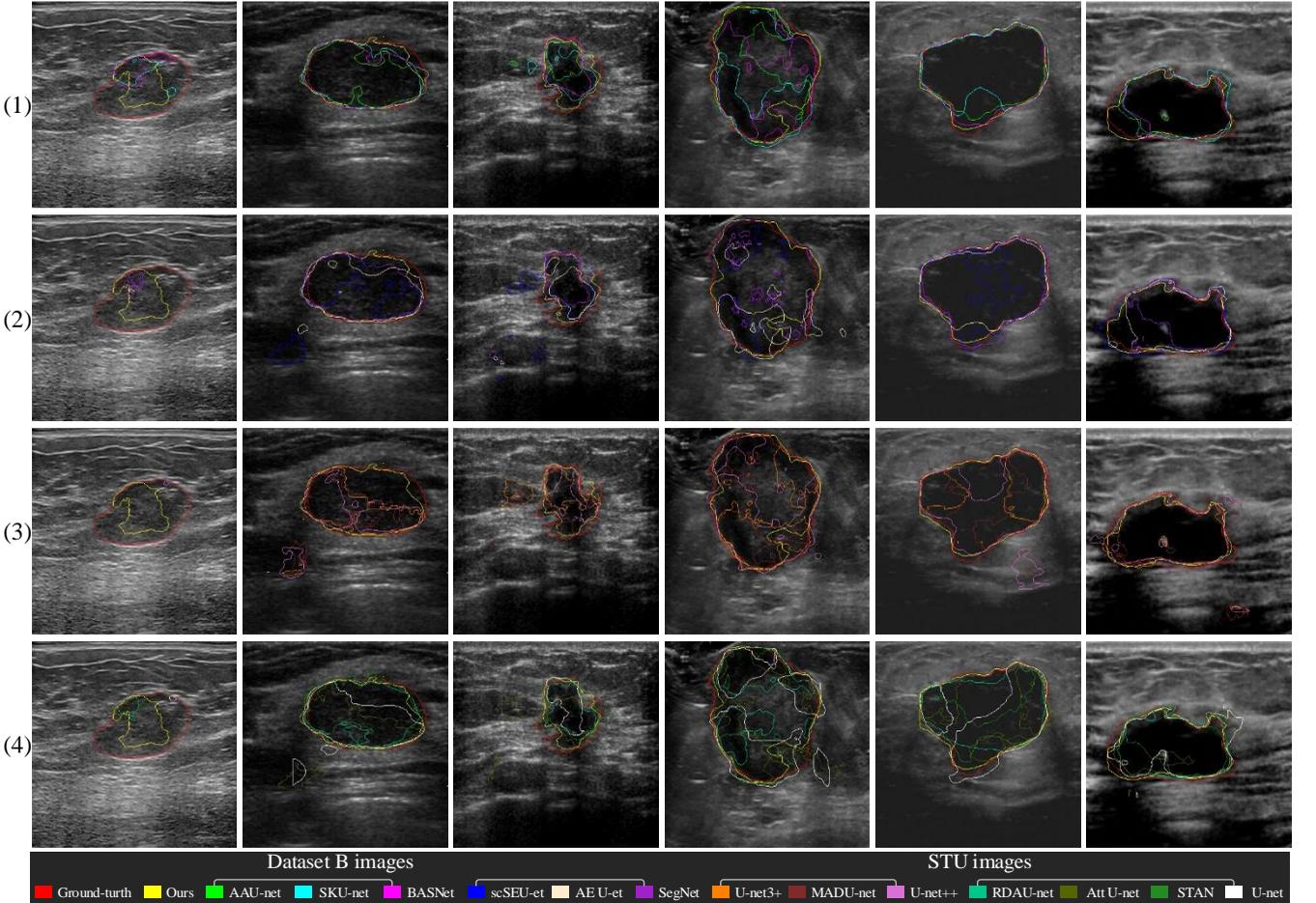

(1)

(2)

(3)

(4)

Dataset B images          STU images

| 🟥 Ground-truth | 🟨 Ours | 🟩 AAU-net | 🟦 SKU-net | 🟪 BASNet | 🟨 scSEU-et | ⬜ AE U-et | 🟩 SegNet | 🟧 U-net3+ | 🟫 MADU-net | 🟪 U-net++ | 🟪 RDAU-net | 🟩 Att U-net | 🟦 STAN | ⬜ U-net |

Fig. 7. The qualitative results of all methods on two external datasets. For visible comparison, we divide the competing methods into four groups, and each group simultaneously exhibits three or four competing methods, the proposed method and the ground-truth contour. (1) AAU-net, SKU-net, BASNet, (2) scSEU-net, AE U-net, SegNet, (3) U-net3+, MADU-net, U-net++, and (4) RDAU-net, Att U-net, STAN, U-net.

Fig. 8 that our method can effectively alleviate the disturbance of uneven energy distribution, blurred boundaries and similar surrounding tissue to the segmentation results. The effectiveness of the method is further verified by segmentation of renal ultrasound images.

TABLE 7
Comparison with Different Methods on Renal Ultrasound Images (mean±std). ASTERISKS INDICATE THAT THE DIFFERENCE BETWEEN OUR METHOD AND THE COMPETING METHOD IS SIGNIFICANT USING A PAIRED STUDENT'S T-TEST. (*: P < 0.05)

| Methods | Jaccard | Precision | Recall | Specificity | Dice |
|---|---|---|---|---|---|
| U-net | 85.00±1.39 | 93.03±1.86 | 91.11±0.79 | 98.77±0.60 | 91.57±0.86 |
| STAN | 86.01±0.48 | 92.64±1.19 | 92.51±0.65 | 98.79±0.42 | 92.12±0.32 |
| Att U-net | 85.99±1.50 | 91.56±1.99 | 93.70±0.32* | 98.50±0.46 | 92.10±0.92 |
| RDAU-net | 83.13±1.02 | 93.53±1.32 | 85.80±2.60 | 98.97±0.50 | 90.47±0.56 |
| U-net++ | 85.59±1.53 | 90.33±1.69 | 93.22±0.22 | 98.28±0.51 | 91.78±1.01 |
| MADU-net | 86.16±1.66 | 93.24±1.46 | 91.54±2.54 | 98.95±0.49 | 92.27±1.09 |
| U-net3+ | 86.35±1.41 | 91.73±2.07 | 93.33±0.51 | 98.52±0.49 | 92.32±0.80 |
| SegNet | 85.78±0.98 | 92.66±1.47 | 92.31±0.89 | 98.67±0.63 | 91.90±0.60 |
| AE U-net | 83.90±4.03 | 91.98±1.22 | 90.91±5.88 | 98.54±0.44 | 90.79±2.61 |
| scSEU-net | 51.36±4.04 | 70.38±5.40 | 62.34±3.93 | 98.85±0.67 | 61.89±4.61 |
| BASNet | 88.42±0.50* | 94.28±1.55 | 93.51±0.89 | 98.94±0.40 | 93.65±0.32* |
| SKU-net | 87.16±0.42 | 93.23±1.36 | 93.36±0.93 | 98.72±0.37 | 92.91±0.25 |
| AAU-net | 86.75±0.86 | 93.67±1.59 | 92.50±0.70 | 98.76±0.49 | 92.64±0.56 |
| C-Net | 88.30±1.39 | 94.30±0.81* | 93.60±1.36 | 99.18±0.16* | 93.51±0.97 |
| Ours | **89.02±1.09** | **94.69±0.73** | **93.95±0.78** | **99.27±0.45** | **94.05±0.66** |

# 7 DISCUSSIONS

## 7.1 Comparison with U-net Variant Network

In this study, we propose an unpretentious nested U-net (NU-net) by utilizing U-nets with different depths to refine the encoded features. Compared with the existing U-net variant network, this method increases the network parameters without significantly increasing the computational cost, as shown in Fig. 9. Furthermore, the extra operations (MOU and MDSC modules) introduced in our approach are very easy to understand and perform.

TABLE 4 and Fig. 10 shows the segmentation results of different segmentation methods on BUSI and Dataset B with four-fold cross-validation. Obviously, our method achieves state-of-the-art results on five quantitative evaluation metrics. As shown in Fig. 10, our method can effectively reduce the missed detection rate and false detection rate and achieve the segmentation results that are closer to the ground-truth mask. In addition, our method can effectively alleviate the perturbation of heterogeneous, blurred boundary and tumor morphology on the segmentation results, as shown in Fig. 10. According to the segmentation results of U-net and U-net variant network two key points can be observed: (1) The existing U-net variant network is more sensitive and less robust to different breast ultrasound images. (2) Although these variant networks improve the segmentation accuracy of breast tumors, there are serious missed detections and false detections on individual images, and even the region of interest cannot be detected.

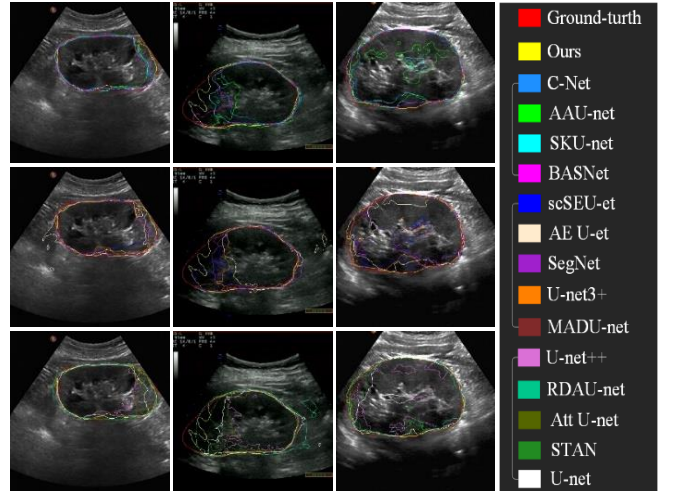

Fig. 8. The segmentation results of different method on renal ultrasound images. We divide the competing methods into three groups, and each group simultaneously exhibits four or five competing methods, our method and the ground-truth contour. (1) C-Net, AAU-net, SKU-net, BASNet, (2) scSEU-net, AE U-net, SegNet, U-net3+, MADU-net, and (3) U-net++, RDAU-net, Att U-net, STAN, U-net.

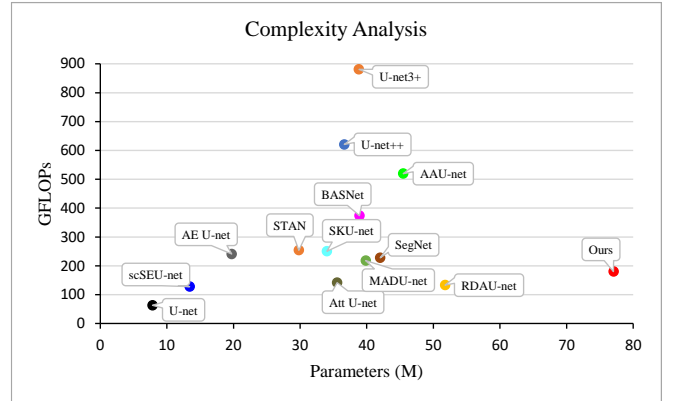

Fig. 9. The complexity analysis of different methods. Obviously, our method has more network parameters, but does not significantly improve the computational cost.

## 7.2 Robustness Analysis

As shown in TABLE 5, our method achieves the best segmentation accuracy on benign tumors and malignant tumors, which indicates that the method has better adaptability to different classes of breast tumors. From TABLE 5 we can see that STAN, Att U-net and RDAU-net reduce the segmentation accuracy of malignant tumors compared to U-net, which indicates that they are more sensitive to complex malignant ultrasound images. The remaining variant networks still improve the segmentation accuracy of breast tumors (benign or malignant). The failure rate of each method for benign and malignant tumors segmentation also further illustrates the robustness of our method, as shown in Fig. 6. Our method achieves minimal segmentation failure rates on benign and malignant tumors. Although, U-net and scSEU-net also achieved the smallest failure rate on malignant tumors segmentation, the overall segmentation performed poorly on the malignant tumors. Similarly, the segmentation performance of scSEU-net and SKU-net on the benign tumors is also the same.

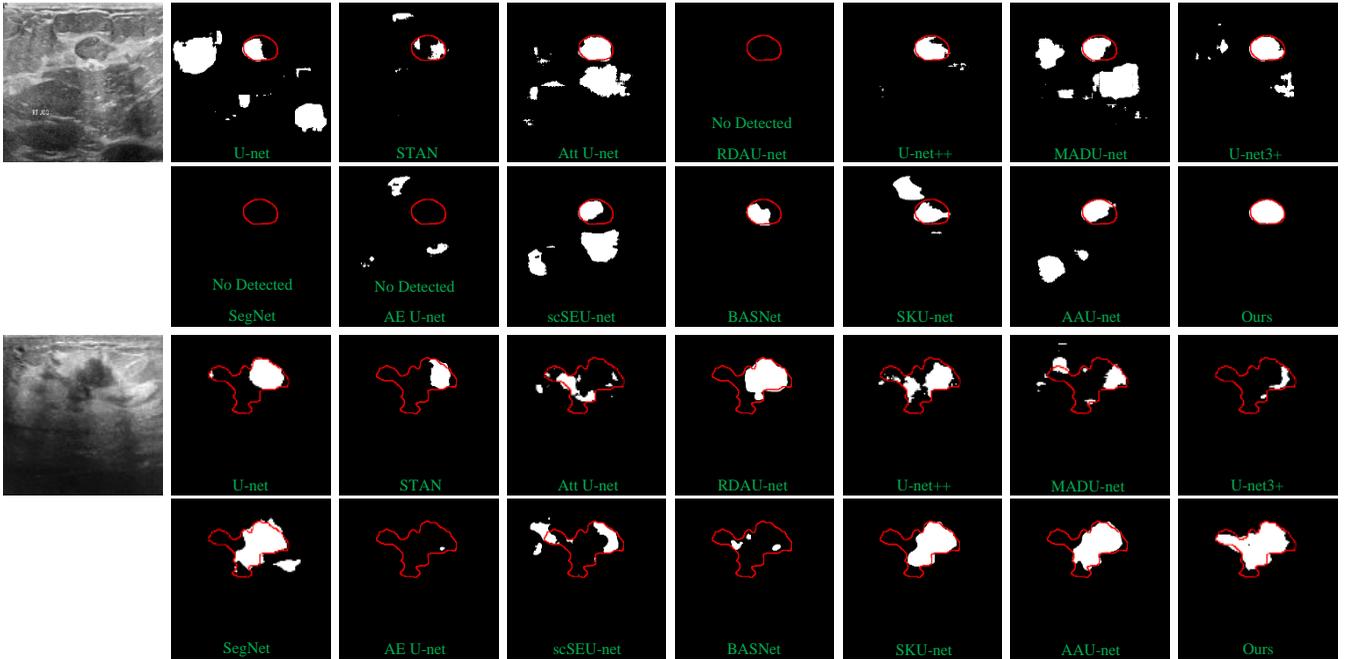

Fig. 10. The segmentation results of different method on BUSI and Dataset B. Some methods fail to segment breast tumors on individual ultrasound images.

The robustness of our method for segmenting breast tumors is further demonstrated through externally validated experiments. From TABLE 6, we can see that AA U-net, SKNet and BASNet still obtain competitive results among the compared methods, which indicates that these three methods have certain potential in breast lesions segmentation. scSEU-net achieves good results in the external experiments of BUSI, but it does not perform well in the external experiments of Dataset B. This indicates that the method is sensitive to different breast ultrasound images and has poor robustness. Similarly, the remaining methods suffer from the same problem. In addition, it can be clearly seen from Fig. 7 that various interference factors cause serious missed detections and false detections in the segmentation results, and even the segmentation fails. Our method can mitigate the interference of these factors and obtain segmentation results that are closer to the ground-truth masks.

### 7.3 Satatistical Test

To investigate the statistical significance of the developed network over compared methods on different quantitative metrics, we perform statistical analysis of p-values in each comparative experiment. As shown in TABLE 4, TABLE 5, TABLE 6 and TABLE 7, we can find that the p-values are almost smaller than 0.05 for the five metrics, demonstrating that our method can be regarded as reaching a significant improvement over the other compared methods on these evaluation metrics. Overall, the method proposed in this paper has good robustness and generalization ability and is more suitable for the segmentation of breast tumors.

## 8 Conclusions

Through the analysis of U-net and its variant networks,

we discover some of their limitations: (1) Tending to use shallower U-nets. (2) Introducing complex extra operations. (3) Reproducing and applying is inconvenient. To alleviate these limitations and further improve the segmentation accuracy of breast tumors, we propose a simple nested U-net (NU-net). The design of this architecture not only reduces the sensitivity of the network to input images with different scales, but also further improves the ability to characterize object or region features. Extensive experiments (comparative experiments, robustness analysis and external validation) with several state-of-the-art deep learning segmentation methods demonstrate that our method has better performance on breast lesions segmentation. In addition, the application on renal ultrasound images also demonstrates the robustness of our method.

## Acknowledgments

This work is supported by the National Natural Science Foundation of China (grant number: U1913207, 51875394).

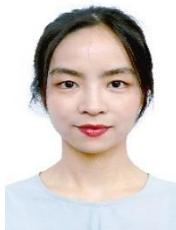

**Lei Li** is a post-doc in the Institute of Biomedical Engineering at University of Oxford. She obtained her PhD degree in 2021 from the School of Biomedical Engineering, Shanghai Jiao Tong University, supervised by Professor Xiahai Zhuang. She visited King's College London in 2020 as a joint PhD student and was supervised by Professor Julia A Schnabel. Her research interest is medical image analysis.

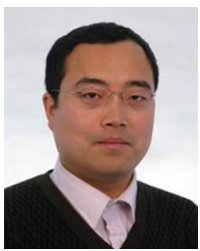

**Yu Dai** received the M.S. and Ph.D. degrees in instrument science and technology from the Harbin Institute of Technology, in 2004 and 2009, respectively. He is now a professor at Nankai University, leading the Medical & Robotics Lab. His research interests include deep learning, medical image analysis and signal processing. He is a member of the IEEE.

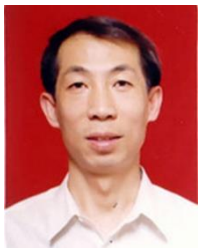

**Jian-xun Zhang** received the M.S. degree in automatic control from Tianjin University, in 1987, and the Ph.D. degree in automatic control from the Institute of Automation, Chinese Academy of Sciences, in 1994. He is now a professor at Nankai University, leading the Medical & Robotics Lab. His research interests include medical robots and robot control. He is a member of the IEEE.

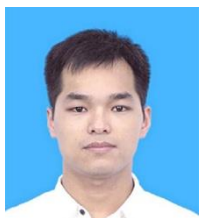

**Gong-ping Chen** is a Ph.D. student from the College of Artificial Intelligence, Nankai University. His research interests include medical image analysis and deep learning, especially for ultrasound image segmentation. He is a graduate student member of the IEEE.